\documentclass{iaus}
\usepackage{graphicx}

\title[Astrophysical MHD simulations] 
{Historical perspective on astrophysical MHD simulations}

\author[Michael L. Norman]   
{Michael L. Norman$^{1,2}$}

\affiliation{$^1$ Center for Astrophysics and Space Sciences
 \\ $^2$ San Diego Supercomputer Center \\ La Jolla, CA 92093, USA
 \\email: {\tt mlnorman@ucsd.edu}}

\pubyear{2010}
\volume{270}  
\pagerange{1-11}
\jname{Computational Star Formation}
\editors{J. Alves, B. G. Elmegreen, J. M. Girart, V. Trimble, eds.}
\begin{document}

\maketitle

\begin{abstract}
This contribution contains the introductory remarks that I presented at IAU Symposium 270 on ``Computational Star Formation" held in Barcelona, Spain, May 31 -- June 4, 2010. I discuss the historical development of numerical MHD methods in astrophysics from a personal perspective. The recent advent of robust, higher order-accurate MHD algorithms and adaptive mesh refinement numerical simulations promises to greatly improve our understanding of the role of magnetic fields in star formation.
\end{abstract}

\firstsection 
\section{Introduction}

It is a distinct pleasure to be invited to speak to you today about numerical MHD simulations of star formation. Moreover it is a great honor to speak second following Richard Larson, whom I consider the founder of computational star formation. As I will relate, his research influenced me in ways he is probably unaware of, and it is nice to have the opportunity to tell that story. I must admit this is the first historical perspectives talk I have been asked to give which means I must be getting old. On the other hand I cannot deny that I have been meddling in computational star formation on and off for 35 years now and have a few reminiscences and battle scars to relate. In this short contribution I do not attempt to be comprehensive about the given topic, but rather describe my personal experiences developing and applying numerical MHD methods to problems of interest, including star formation.
 
\section{Caltech coincidences}
Before I do that I must relate a couple of strange coincidences that occurred to me when I was an undergraduate at Caltech which in hindsight foreshadowed my graduate research at Livermore. First, as a new freshman I wandered into Millikan Library--a Caltech landmark--to browse the astronomy and physics library. I saw a shelf filled with beautifully bound red volumes, and picked one off the shelf at random to see what they were. I picked Richard Larson's PhD thesis which I would later, as a graduate student, study in great detail. At the time though I didn't understand anything and could barely comprehend how a PhD thesis came into existence. I flipped through it, impressed with the graphs and equations, and put it back on the shelf. The second foreshadowing occurred when I was a sophomore or junior. I did a term paper on supernova explosions for Peter Goldreich's class on the interstellar medium. In the process I ran across a paper in the Astrophysical Journal written by Jim Wilson on numerical simulations of neutrino-driven iron core collapse supernova explosions. Jim would later become my PhD thesis advisor and suggest a topic in star formation that would eventually bring me into contact with Richard Larson's early research.
 
\section{Livermore Years}
I did my PhD thesis on numerical star formation under the supervision of Jim Wilson at the Lawrence Livermore National Laboratory from 1975 to 1980. Jim was one of the true pioneers of numerical astrophysics (\cite{Centrella85}), and I was fortunate to have him as my supervisor. He was absolutely fearless when it came to tackling a new problem numerically. This was due to the fact that in the 1960s he had developed 2D multiphysics codes to simulate the internal operations of nuclear weapons, which gave him an encyclopedic knowledge of hydrodynamics and MHD, neutronics and radiative transfer, plasma physics, nuclear reactions, etc. In the late 60s Jim became interested in astrophysics and started to work on core collapse supernovae, relativistic stars, magneto-rotationally driven jets, and somewhat later, numerical general relativity. In the 1970s Jim had assisted David Black and Peter Bodenheimer at UC Santa Cruz develop a 2D hydro code which they applied to axisymmetric rotating protostellar cloud collapse simulations (\cite{BB75,BB76}). They found the collapse produced a gravitationally bound ring, confirming a result published by Richard Larson in 1972. Jim suggested I look at the stability of this ring to nonaxisymmetric perturbations using a 3D self-gravitating hydro code he had written. I said OK. He gave me two boxes of IBM punch cards and said get to work. I did, and two years later I had my first publication (\cite{Norman78}).
 
For my PhD thesis I developed a new 2D axisymmetric Eulerian hydro code to study rotating protostellar cloud collapse. Years later this code would become the basis for the first ZEUS code. I showed that the self-gravitating ring seen by Larson (1972) and Black \& Bodenheimer (1976) was a numerical artifact produced by spurious transport of angular momentum (\cite{Norman80}). I presented this result, and the truncation error analysis it was based on, at the 1979 Santa Cruz star formation summer school. Larson, Black, and Bodenheimer were in the audience. Here I was, an unknown graduate student, telling the big names in the field that their results were incorrect in front of the star formation community. Afterwards Richard was very gracious about it. 

That work taught me an important lesson about numerical simulations which I have never forgotten and young researchers should not forget: that numerical errors masquerade as physics, and that one needs to not take numerical results at face value. A high level of skepticism needs to be applied to any new and interesting result, because it may simply be wrong. The code may simply be doing the best it can under difficult circumstances. Numerical star formation, with its vast range of scales, is a very difficult problem. This I learned reading Richard Larson's thesis.  

\section{Protostars and Planets, Tucson, 1978}
I became aware of the importance of magnetic fields to star formation when I attended the first Protostars and Planets meeting in Tucson, Arizona in January 1978. That is where I met Richard Larson for the first time. All the big names were there, including George Field, Hannes Alfv\'{e}n, and Joe Silk. Chaisson and Vrba talked about magnetic field structures in dark clouds. Field talked about conditions in collapsing clouds, and John Scalo talked about the stellar mass spectrum. A combative young astrophysicist by the name of Telemachos Mouschovias presented theoretical models of magnetically supported clouds, and how ambipolar diffusion would lead to gravitational instability once a critical mass to magnetic flux ratio was exceeded. This work is exceedingly well known now, but in 1978 it was still rather new. One of my strongest recollections of the conference was the Q \& A after Alfv\'{e}n's talk. Mouschovias and Alfv\'{e}n were in violent agreement about the fundamental importance of magnetic fields to star formation, but seemed to agree on nothing else. That evening I presented a 16mm movie of my 3D hydrodynamic ring fragmentation instability simulations to a receptive audience. But by then I was convinced I was solving the wrong equations, and that what was really required was 3D MHD simulations with ambipolar diffusion and self-gravity—a tall order. If fact, this was what Jim Wilson suggested I work on for my thesis, but I got side-tracked on the 2D axisymmetric work and then decided it was time to graduate. Nonetheless, the takeaway that astrophysical fluid dynamics is fundamentally MHD, not HD, was strongly impressed on me.

Fig. 1 summarizes the current view of the star formation process, and the role magnetic fields are thought to play. We tend to organize the subject around objects at different length scales, proceeding from the largest (giant molecular clouds or complexes) to the smallest (protostars). In between are clouds, cloud cores, protostellar accretion disks and jets. Magnetic fields appear to be important at all these scales, and at some scales fundamental. In the last column I list the minimum useful computational model to study these objects. The importance of magnetic fields seems to increase with decreasing length scale, while to difficulty of numerical modeling increases with increasing length scale because of the lack of simplifying symmetries.

\begin{figure}[t]
\begin{center}
 \includegraphics[width=\textwidth]{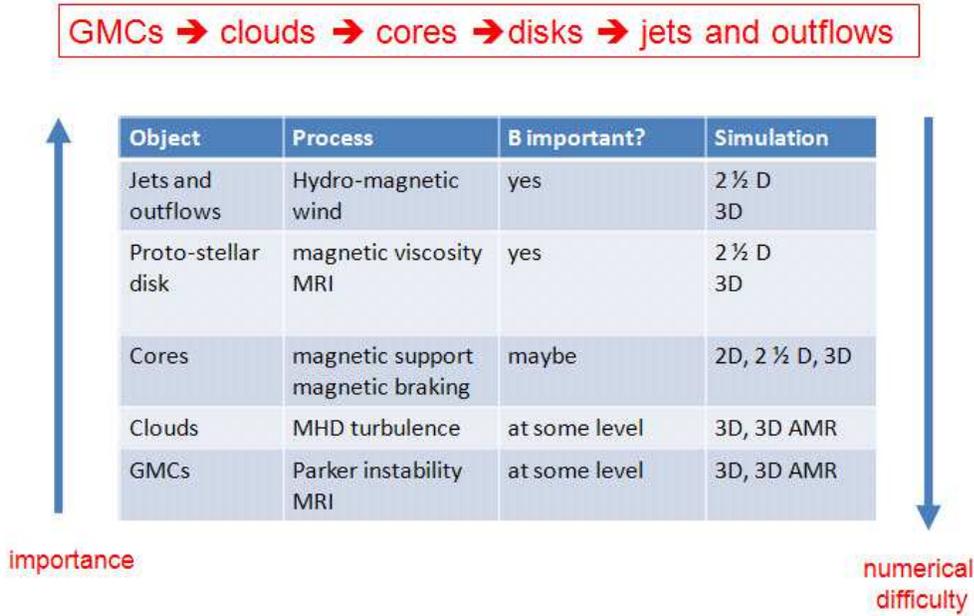} 
 \caption{ Magnetic fields and star formation. We understand star formation as a sequence of related objects and phenomena involving self-gravity, magnetic fields, and turbulence. The importance of magnetic fields seems to increase with decreasing length scale, while to difficulty of numerical modeling the relevant systems increases with increasing length scale because of the lack of simplifying symmetries.}
   \label{fig1}
\end{center}
\end{figure}

\section{Astrophysical Jets}
After graduation, my career took a decade-long detour into simulations of astrophysical jets. It was this application, not protostars that got me seriously and permanently involved in developing numerical MHD methods. The VLA had just come online and was producing spectacular radio maps of extragalactic radio jets like Cygnus A which were undeniably magnetized. Hydromagnetic launching mechanisms were being proposed by Blandford \& Payne (1982) for radio jets, and by Pudritz \& Norman (Colin) (1986) and Shibata \& Uchida (1985) for protostellar jets. My first simulations of radio jets were purely hydrodynamic, carried out with an improved version of my thesis code. But by 1986 I had incorporated magnetic fields. Working with University of New Mexico radio astronomer Jack Burns and his graduate student David Clarke, we applied this code to magnetically-confined supersonic jet models of extragalactic radio sources (\cite{CNB86}).

\begin{figure}[t]
\begin{center}
 \includegraphics[width=\textwidth]{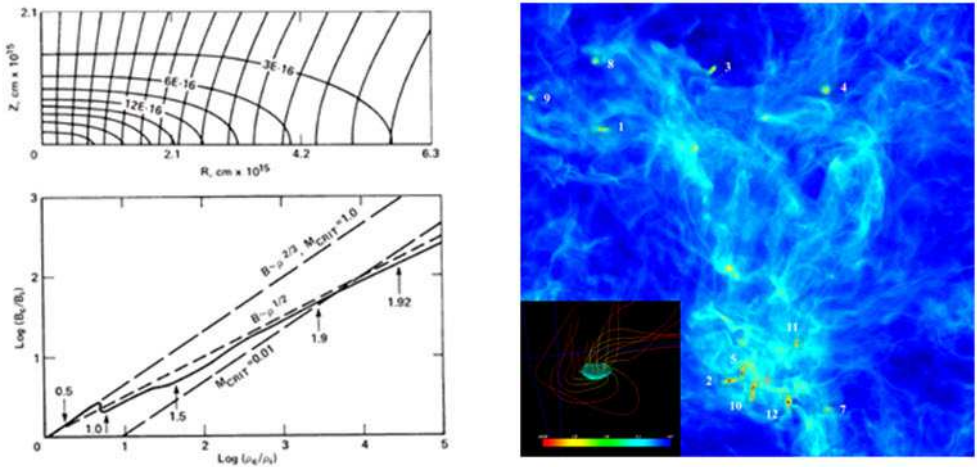} 
 \caption{ Progress with MHD simulations of star formation. Left: flattened cloud core and central B-rho relation in a 2D non-rotating magnetized collapse simulation (from \cite{SB80}). Right: self-gravitating cores in a 3D simulation of super-Alfv\'{e}nic turbulence. Inset: magnetic field topology in a core (from \cite{Li_etal04}).}
   \label{fig2}
\end{center}
\end{figure}

\section{Evolution of Numerical MHD}
\subsection{Early Days}
The development and application of numerical MHD to problems in star formation lagged HD simulations by over a decade because the simplest nontrivial problem is 2D axisymmetric, where as the early hydrodynamic work could be done in 1D spherical symmetry (e.g., Larson 1969, Westbrook \& Tarter 1975). Although Mouschovias had already published by the mid 1970s 2D static models of magnetically supported clouds, it was not until 1980 that the first dynamic MHD simulation was published. Scott \& Black (1980) simulated the gravitational collapse of a non-rotating cloud threaded by a uniform magnetic field. They used a first order upwind scheme (donor cell) to evolve the poloidal flux function, ensuring divergence-free poloidal fields. They showed that collapse produces flattened cores as expected, and that the central density and magnetic field scale as $B_c \propto \rho_c^{1/2}$ (Fig. 2a).

Motivated by the recently discovered jets from young stellar objects, Shibata \& Uchida (1985) carried out 2-1/2D axisymmetric MHD simulations of hydromagnetically-driven disk wind models. The difference between a 2D and a 2-1/2D simulation is that in rotating axisymmetric systems, toroidal velocity and magnetic components are also evolved. Their so-named "sweeping magnetic twist" mechanism rediscovered much earlier work by LeBlanc \& Wilson (1970) in which rotation efficiently coverts poloidal B-fields into toroidal B-fields, producing what is in effect a coiled magnetic spring that uncoils along the rotation axis due to magnetic pressure, launching a jet. They evolved all three components of B using the second order Lax-Wendroff method stabilized with artificial viscosity. Such an approach is not guaranteed to maintain divergence-free B-fields.

Clarke, Burns \& Norman (1989) performed 2-1/2D MHD simulations of extragalactic radio jets using the original code called ZEUS. The code evolved the poloidal flux function and the toroidal component of the magnetic field using 2nd-order upwind finite differences. This ensures divergence-free magnetic fields as can easily be demonstrated. The poloidal flux function is defined $a_\phi = rA_\phi$, where $r$ is the cylindrical radius and $A_\phi$ is the magnetic vector potential. We then have $B_r=-\frac{1}{r}\frac{\partial a_\phi}{\partial z}, 
B_z=\frac{1}{r}\frac{\partial a_\phi}{\partial r}$.
By virtue of the axisymmetry of the toroidal field $B_\phi$ it is evident that 
\[
\nabla \cdot \vec{B}=\frac{\partial B_z}{\partial z}
+\frac{1}{r}\frac{\partial rB_r}{\partial r}
+\frac{1}{r}\frac{\partial B_\phi}{\partial \phi}=
+\frac{1}{r}\frac{\partial^2 a_\phi}{\partial z \partial r}
-\frac{1}{r}\frac{\partial^2 a_\phi}{\partial r \partial z}
+0=0. 
\]
Faraday's law for evolving the magnetic field becomes
\[
\frac{\partial B_\phi}{\partial t}
+\frac{\partial}{\partial r}(B_\phi v_r)
+\frac{\partial}{\partial z}(B_\phi v_z)
=r \vec{B} \cdot \nabla \Omega 
\]
\[
\frac{\partial a_\phi}{\partial t}
+v_r \frac{\partial a_\phi}{\partial r}
+v_z \frac{\partial a_\phi}{\partial z}=0,
\]

\noindent
where $\Omega = v_\phi /r$. These equations were evolved in ZEUS using a second-order monotonic upwind scheme alongside the hydrodynamic equations, with the Lorentz force term constructed from first and second difference of $B_\phi$ and $A_\phi$. This was a very neat, stable, and reasonably accurate scheme for 2-1/2D MHD simulations. However it could not be generalized to 3D, and therefore a divergence-free method working directly with the components of B had to be found.

\subsection{Constrained Transport}
Fortunately, in 1988 Evans \& Hawley solved half the problem when they introduced the Constrained Transport (CT) method. CT solves the magnetic induction equation in integral form and uses a particular centering of the magnetic and velocity field components in the unit cell so as to transport vector B through a 3D mesh in a divergence-free way. For a recent exposition of this see Hayes et al. (2006). I say they only solved half the problem because what they addressed was how to treat the kinematics of magnetic fields, not their dynamics. As we discuss below, an accurate and stable treatment of the dynamics of magnetic fields requires judicious choices for how the EMFs and Lorentz force terms are evaluated.
 
\subsection{ZEUS and Sons}
In 1987 University of Illinois grad student Jim Stone and I set out to build a version of the Clarke-Norman ZEUS code that evolved ($B_z, B_r, B_\phi$) in a divergence-free way using CT (Evans visited NCSA in 1987 and told us about it). We figured if we could make this work in 2-1/2D, it could easily be generalized to 3D. The end result of this effort was a code called ZEUS-2D (Stone \& Norman 1992a,b), developed by Jim, and a code called ZEUS-3D, developed by Clarke who became my postdoc in 1988. Jim and I were motivated to improve on the hydromagnetic disk wind simulations of Shibata \& Uchida (1985). When we tried CT as described by Evans \& Hawley (1988) it failed miserably to stably evolve the torsional Alfv\'{e}n waves generated when the rotating disk starts twisting the initial poloidal field (see Fig. 3). The reason for this is that the EMFs used in the vanilla CT scheme were not upwind in the Alfv\'{e}n wave characteristics, but rather were computed using simple centered differences and averages. Jim and I came up with a different way to calculate the EMFs using a Method of Characteristics approach specifically for Alfv\'{e}n waves. The resulting hybrid scheme we called MOC-CT. It worked beautifully on the torsional Alfv\'{e}n wave problem (Fig. 3b) and 2-1/2D simulations magnetized accretion disks (Stone \& Norman 1994).

Using ZEUS-2D, we accidently discovered the MRI in 1989 (Norman \& Stone 1990) but didn't realize the significance of what we were seeing. Several years later, with the pioneering work of Balbus and Hawley we realized what we had computed was the axisymmetric channel solution of the MRI (Stone \& Norman 1994). This anecdote is the counter-example to what I said above about being skeptical of numerical results. In this case, the simulations contained a discovery which we failed to recognize. It was present in the simulation because we had improved the algorithm to the point where we didn't need excessive amounts of artificial viscosity to damp numerical instabilities. 

\begin{figure}[t]
\begin{center}
 \includegraphics[width=\textwidth]{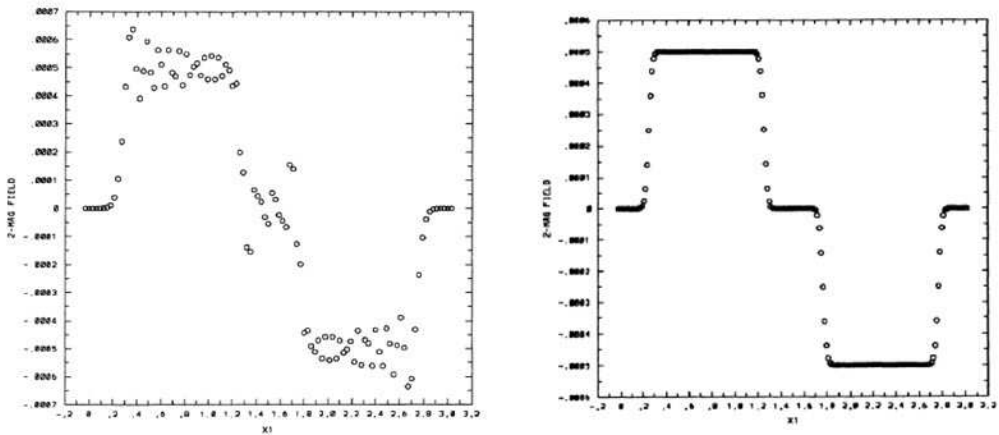} 
 \caption{Importance of upwinding. Shear Alfv\'{e}n wave test problem. {\em Left:} computed using original CT algorithm of Evans \& Hawley (1988);{\em Right:} computed using MOC-CT algorithm of Stone \& Norman (1992b). From Stone \& Norman (1992b).}
   \label{fig3}
\end{center}
\end{figure}

In 1989, David Clarke began a 3D implementation of MOC-CT algorithm which resulted in the ZEUS-3D code. However he encountered explosive numerical instabilities when he applied it to magnetized extragalactic jet models. The lobes of these jets exist in a state of super-Alfv\'{e}nic turbulence, which as we later learned from simulations of molecular cloud turbulence, is very tough on numerical schemes. Hawley and Stone eliminated the numerical instability using the tried and true method of adding numerical dissipation (Hawley \& Stone 1995). This fix was incorporated into ZEUS-3D, and at that point it was publicly released via the Laboratory for Computational Astrophysics (LCA) website.
 
Subsequently, ZEUS-3D became widely used for many different kinds of applications including some of the earliest work on decay rates in turbulent molecular clouds (MacLow et al. 1998). The stabilized MOC-CT of Hawley and Stone made its way into other code implementations by Hawley, Stone, Gammie, Eve Ostriker, and others who have done important work on the MRI, molecular cloud turbulence, protostellar accretion disks, and Galactic ISM dynamics. The LCA developed its own MPI-parallel version of ZEUS-3D called ZEUS-MP (Norman 2000) which is now in its version 2.0 release (Hayes et al. 2006). Fig. 4 shows the ``ZEUS diaspora" to the best of my knowledge.
 
\begin{figure}[t]
\begin{center}
 \includegraphics[width=\textwidth]{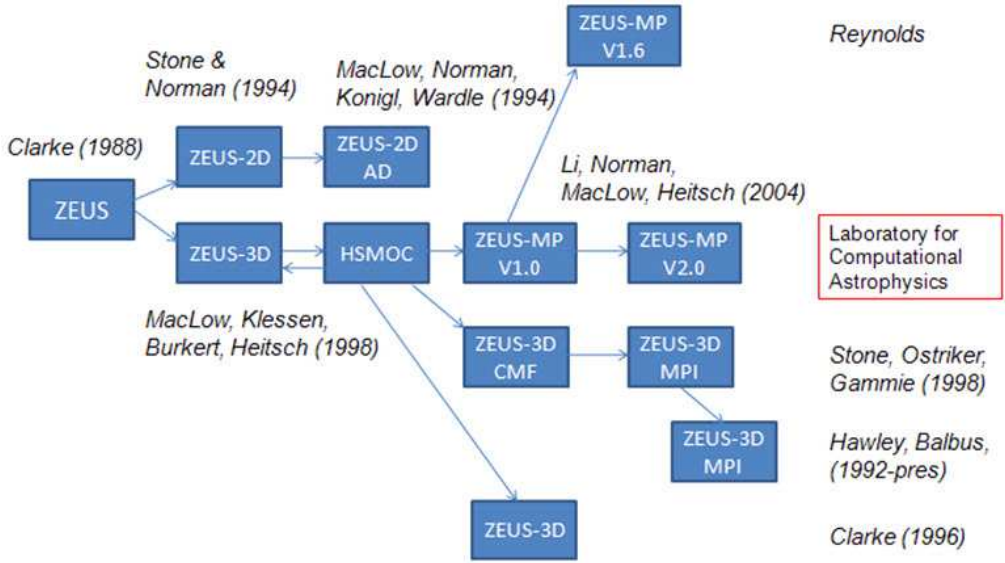} 
 \caption{The ``ZEUS diaspora". ZEUS' MHD algorithms have made their way into a number of code implementations. A few citations to significant contributions to computational star formation are included.}
   \label{fig4}
\end{center}
\end{figure}

To cite just a few significant applications of ZEUS to computational star formation, I would mention MacLow et al. (1998) and Stone, Ostriker \& Gammie (1998) on turbulence decay rates in molecular clouds, Heitsch, MacLow \& Klessen (2001) on self-gravitating molecular cloud turbulence, and Gammie et al. (2003) and Li et al. (2004) on gravitationally bound core formation in turbulent molecular clouds, and MacLow et al. (1995) incorporating ambipolar diffusion into ZEUS. 

\subsection{A solar physicist gets involved}
The field of computational star formation was enlivened when Paolo Padoan went to Copenhagen for his PhD research in the mid-90s. There he joined forces with $\AA$ke Nordlund, a prominent solar physicist who not surprisingly was in possession of a 3D compressible MHD code called the STAGGER code. The codes solved the ideal MHD equations in non-conservative form on a 3D staggered mesh of size 128$^3$, higher-order finite differences, and stabilized using artificial viscosity. The divergence-free condition on the magnetic field was not enforced. Using this code they carried out isothermal compressible MHD turbulence-in-a-box simulations ignoring gravity and other effects. Varying the Alfv\'{e}n Mach number, they showed that a super-Alfv\'{e}nic model more closely match a variety of molecular cloud observations than a trans-Alfv\'{e}nic model (Padoan \& Nordlund 1999). This set the stage for their turbulent fragmentation theory of star formation (Padoan \& Nordlund 2002) which has been very influential in the field. They argued that there is a direct link between the mass function of gravitationally bound cores and the statistical properties of super-Alfv\'{e}nic turbulence. Pakshing Li, Mordecai Mac-Low, Fabian Heitsch and myself verified this claim with the first 512$^3$ simulation of self-gravitating, super-Alfv\'{e}nic MHD turbulence using ZEUS-MP (Li et al. 2004).
 
\subsection{The Rise of Upwind Schemes}
The last 20 years have witnessed a lot of algorithmic development activity in what is generically called higher order-accurate upwind schemes (or Godunov schemes) for ideal MHD. While details differ, the basic idea is to write the ideal MHD equations in fully conservative form:
\[
\frac{\partial \vec{U}}{\partial t} + \nabla \cdot \vec{F}(\vec{U})=0,
\]
where U is the vector of unknowns, and F is the flux vector, which is a complicated non-linear function of U. For 3D ideal MHD
\[
U^T=(\rho, \rho v_x, \rho v_y, \rho v_z, \mathcal{E}, B_x, B_y, B_z)
\]
where the symbols have their usual meanings and $\mathcal{E}$ is the total energy. Schemes use the divergence theorem to update U in the control volume cells by differencing F on the faces. The entire burden and benefit of upwind schemes is to find accurate and stable representations for F that are upwind in all the wave characteristics of MHD. This is accomplished through the use of Riemann solvers, both exact and approximate, of which there are many available. Modern MHD codes implementing upwind schemes are built in a modular fashion, mixing and matching half a dozen basic ingredients in different ways. Within the conservation law solver, these are: order accuracy of the spatial interpolation, order of accuracy of the temporal integration, choice of Riemann solver, choice of monotonicity-preserving flux limiters, directional splitting versus unsplit. Three basic approaches to maintaining the divergence free condition are: 1) ignore it; 2) clean it (elliptic, hyperbolic); 3) prevent it (constrained transport). Varying these choices leads to hundreds of potential combinations, not all of which have been explored. In the following I give a very brief survey of existing methods. 

Zachary \& Colella (1992) developed an exact solver for the MHD Riemann problem. Ryu et al. (1995, 1998) developed 2D and 3D ideal MHD codes based on the Total Variation Diminshing (TVD) method. Dai \& Woodward (1994, 1998) generalized the Piecewise Parabolic Method (PPM) to ideal MHD. Balsara (1998a,b) developed a linearized Riemann solver and improved TVD schemes for adiabatic and isothermal MHD. Balsara \& Spicer (1999) incorporated Constrained Transport (CT) into a TVD MHD scheme. Powell et al. (1999) introduced elliptic divergence cleaning. Londrillo \& del Zanna (2000) introduced further improvements to a TVD MHD scheme. Dedner et al. (2002) introduced hyperbolic divergence cleaning.  Gardner \& Stone (2005) developed an improved PPM+CT scheme and stressed the importance of directionally unsplit schemes. Popov \& Ustyugov (2008) introduce a PPM on a local stencil (PPML) and married it to CT. Ustyugov et al. (2009) added stability improvements to PPML+CT and demonstrated its application to super-Alfv\'{e}nic turbulence.
 
\subsection{AMR MHD}
The severe dynamic range requirements to resolve gravitational collapse in multi-dimensions requires nested grids or adaptive mesh refinement (AMR) to attack. In pioneering work, Dorfi (1982) first introduced static hierarchically refined grids into calculations of isolated cloud collapse including magnetic fields, self-gravity and rotation. Using a simple finite difference scheme he showed that magnetic breaking is 10 times as efficient in the perpendicular rotator case as in the aligned rotator case. Be found that bar-like structures are produced by the collapse and breaking. Hierarchical nested grid codes were also developed by Ziegler \& Yorke (1997), Tomisaka (1998), and Machada et al. (2005, 2005) and applied to isolated core collapse of ever increasing dynamic range. The last authors showed that disk fragmentation is sensitive to magnetic field strength and inclination.

True AMR MHD has only come onto the scene recently due to the numerical challenges involved. It builds on progress made with AMR hydro codes developed by Berger \& Colella (1989), Bryan \& Norman (1997), Truelove et al. (1998), Fryxell et al. (2000), and Teyssier (2002). The first AMR MHD code was the RIEMANN code by Balsara (2001). This was followed by the FLASH code (Linde 2002), the NIRVANA code (Ziegler 2005), the RAMSES code (Fromang et al. 2006), and the ENZO code (Collins et al. 2010a). The last of these, developed by my graduate student David Collins, is the result of a 5 year effort to marry a higher order upwind scheme for ideal MHD with CT on a block structured adaptive mesh. We tried a quite a number of different conservation law solvers, Riemann solvers, and CT strategies before we found one stable enough to deal with the rigors of super-Alfv\'{e}nic turbulence.

\begin{figure}[t]
\begin{center}
 \includegraphics[width=\textwidth]{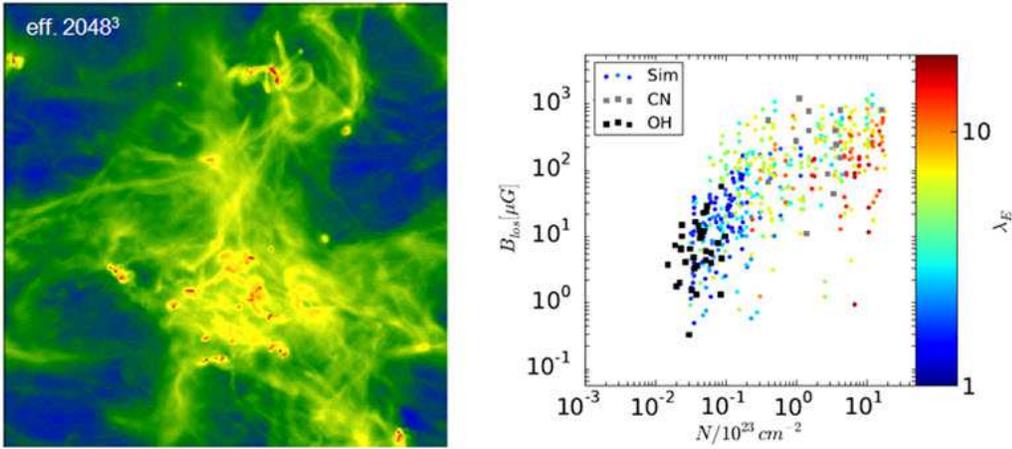} 
 \caption{ AMR MHD simulation of turbulent fragmentation, with an effective resolution of 2048$^3$. {\em Left:} projected gas density in the 10 pc box. {\em Right:} comparison of simulated and observed Zeeman measurements in dense cores. From Collins et al. (2010b).}
   \label{fig5}
\end{center}
\end{figure}

\section{Results, Finally!}
Fig. 5 shows results ENZO-MHD applied to turbulent fragmentation with self-gravity with an effective resolution of 2048$^3$ (Collins et al. 2010b). On the left is projected gas density through the box at 0.75 free-fall times. On the right is a scatter plot of the simulated LOS magnetic field strength versus the gas column density in bound cores overlaid on Zeeman observations. Color coding indicates the ratio of gravitational to magnetic energy in the cores. The simulation is in good agreement with the observations, lending further support to the turbulent fragmentation picture.

\section{Brand New Day}
The AMR MHD simulation shown above was beyond my wildest dreams when I was a graduate student listening to Mouschovias and Alfv\'{e}n square off at Protostars and Planets I. After all these years, I think the field of numerical MHD is finally where we want it to be. With AMR MHD we are finally solving the problem Nature hands us, not some reduced problem (although we have learned a tremendous amount solving reduced problems.) Admittedly, we still have to incorporate ambipolar diffusion, dust, chemistry and cooling, and radiative transfer into AMR simulations. However, the progress being made on all these fronts reported at this meeting are very encouraging. The arsenal of available codes and the number of young people engaged in their development and use encourage me to believe that great progress will continue to be made in computational star formation for years to come. 
\\

{\underline{\it Acknowledgements}}. I gratefully acknowledge my former and current "star formation" students, postdocs, and collaborators from whom I have learned and enjoyed so much and who have kept me challenged and in the game: Tom Abel, Dinshaw Balsara, Greg Bryan, David Clarke, David Collins, John Hayes, Fabian Heitsch, Alexei Kritsuk, Hui Li, Pakshing Li, Shengtai Li, Mordecai MacLow, $\AA$ke Nordlund, Brian O'Shea, Paolo Padoan, Jim Stone, Matt Turk, Sergey Ustyugov, Rick Wagner, and Dan Whalen. Some of this work was partially supported by NSF grant AST-0808184.

\end{document}